\documentclass[aps, superscriptaddress, amsfonts, amssymb, amsmath, twocolumn, floatfix]{revtex4}

\usepackage[USenglish]{babel}
\usepackage{graphicx}
\usepackage{xcolor}
\usepackage{slashed}
\usepackage{siunitx}
\usepackage{hyperref}

\usepackage{soul}
\usepackage[normalem]{ulem}
\usepackage[makeroom]{cancel}

\hypersetup{colorlinks = true, linkcolor = blue, citecolor = blue, filecolor = magenta, urlcolor = cyan, pdfpagemode = FullScreen}

\date{\today}

\begin{document}

\title{Quantum internal vibrations in macroscopic systems with classical centers of mass}

\author{Gabriel H. S. Aguiar}
\email{ghs.aguiar@unesp.br}
\affiliation{Universidade Estadual Paulista, Instituto de F{\'i}sica Te{\'o}rica, Rua Dr. Bento Teobaldo Ferraz 271, Bloco II, 01140-070, S{\~ a}o Paulo, S{\~ a}o Paulo, Brazil}

\author{George E. A. Matsas}
\email{george.matsas@unesp.br}
\affiliation{Universidade Estadual Paulista, Instituto de F{\'i}sica Te{\'o}rica, Rua Dr. Bento Teobaldo Ferraz 271, Bloco II, 01140-070, S{\~ a}o Paulo, S{\~ a}o Paulo, Brazil}


\begin{abstract}
    Harmonizing classical and quantum worlds is a major challenge for modern physics. A significant portion of the scientific community supports the notion that classical mechanics is an effective theory that arises from quantum mechanics. Recently, the present authors have argued that this should not be the case, as quantum mechanics is not trustworthy for describing the center of mass of systems with masses~$m$ much larger than the Planck mass~$M_\text{P}$. In this vein, a simple gravitational self-decoherence model was proposed, describing how the center of mass of quantum systems would classicalize for~$m \sim M_\text{P}$~\cite{2025-aguiar}. Here, we show that our model does not prevent macroscopic systems (with classical centers of mass) from harboring quantum internal vibrations (as has been observed in the laboratory).
\end{abstract}

\maketitle

\section{Introduction}

Matching the quantum and classical laws of the micro and macro realms, respectively, has been of great concern since the conception of quantum mechanics~(QM). The fact that QM governs the elementary blocks of nature hints that macroscopic systems would also be worthy of a quantum description. Those who share this reductionist view credit environmental sources of decoherence for driving macroscopic systems into classical states (see, {\em e.g.}, Refs.~\cite{1989-ellis, 2013-blencowe, 2013-anastopoulos, 2021-asprea, 2022-zurek} and references therein). In this picture, classical mechanics~(CM) would be an effective theory emerging from QM.

A distinct point of view is defended by those who support that QM must be amended in order to explain the classical realm (see, {\em e.g.}, Refs.~\cite {1989-diosi, 1996-penrose, 1998-penrose, 2003-bassi, 2013-bassi, 2017-bassi, 2022-anastopoulos} and reference therein). In this context, CM would be an effective theory emerging from the amended QM.

The present authors favor a third perspective~\cite{2025-aguiar}, according to which QM and CM would be both effective theories emerging from some still unknown fundamental {\em quantum spacetime theory}~(QST). From this point of view, in usual space and time scales ({\em i.e}, much larger than the Planck length~$L_\text{P} \equiv \sqrt{G \hbar / c^3} \sim 10^{- 35} \, \si{\meter}$ and Planck time~$T_\text{P} \equiv \sqrt{G \hbar / c^5} \sim 10^{- 43} \, \si{\second}$), systems with masses~$m \ll M_\text{P}$ and~$m \gg M_\text{P}$ would be described by QM and CM, respectively. The QM and CM realms would be separated by a Heisenberg cut at~$m \sim M_\text{P}$, where new physics would appear. In order to harmonize the~$m \ll M_\text{P}$ and~$m \gg M_\text{P}$ domains, a (Lorentz invariant) gravitational self-decoherence mechanism was proposed, describing how the center of mass~(c.m.) of quantum systems would classicalize as their masses approach the Planck scale. The physical picture is that quantum coherence would effectively flow from the c.m. of the system to the (still unknown) degrees of freedom of some fundamental quantum spacetime. The model ensures that the classicalization process is quite inefficient for masses~$m \ll M_\text{P}$, in contrast to masses at the Planck scale,~$m \sim M_\text{P}$. 

Here, we show that, although the model predicts that the c.m. of macroscopic systems ($m \gg M_\text{P}$) should obey classical rules, it does not prohibit the internal degrees of freedom from behaving in accordance with QM. For this purpose, we consider a composed system with internal-vibration modes and show that those with frequencies~$\omega$ much smaller than the Planck one~$\Omega_\text{P} \equiv 1 / T_\text{P} \sim 10^{43} \, \si{\hertz}$ will effectively behave in accordance with QM (see Ref.~\cite{2023-bild} for an experimental realization). [Considering frequencies much larger than~$\Omega_\text{P}$ would not be trustworthy for the same reason that our present spacetime description (as a smooth manifold endowed with a Lorentzian metric) would not be reliable at those scales either~\cite{2025-aguiar}.]

The paper is organized as follows. In Sec.~\ref{GSD-CM}, we briefly review our simple gravitational self-decoherence model~\cite{2025-aguiar}. In Sec.~\ref{GSD-IVM}, we adapt the model to treat the internal-vibration modes of a composed system. Our conclusions appear in Sec.~\ref{C}. Hereafter,~$G = c = 1$ and all observables are expressed in powers of seconds. In this unit system,~$M_\text{P} = L_\text{P} = T_\text{P} = \hbar^{1 / 2}$.

\section{Gravitational self-decoherence: center of mass}
\label{GSD-CM}

Let us consider a composed system with mass~$m \lesssim M_\text{P}$, where the c.m. degrees of freedom are decoupled from the internal ones, being described by the Hamiltonian
\begin{equation}
    H(\mathbf{r}, t)
    \equiv
    K(\mathbf{r}) + V(\mathbf{r}, t)
\end{equation}
with kinetic term
\begin{equation}
    K(\mathbf{r})
    \equiv
    - [\hbar^2 / (2 m)] \nabla_\mathbf{r}^2
\end{equation}
and external potential~$V(\mathbf{r}, t)$. According to QM, the wavefunction of the c.m. is unitarily evolved by
\begin{equation}
    i \hbar \partial_t \psi(\mathbf{r}, t)
    =
    H(\mathbf{r}, t) \psi(\mathbf{r}, t),
    \label{QM-CM-E}
\end{equation}
preserving information.

In order to describe how the c.m. of a system with a mass at the Planck scale would classicalize, our effective model assumes that the c.m. gravitationally interacts with a (non-observable) {\em virtual clone} of itself. Thus, by preparing the c.m. in a state~$\psi(\mathbf{r}, 0)$, the clone would be driven to~$\psi(\bar{\mathbf{r}}, 0)$, such that the combined wavefunction would be
\begin{equation}
    \Psi(\mathbf{r}, \bar{\mathbf{r}}, 0)
    =
    \psi(\mathbf{r}, 0) \psi(\bar{\mathbf{r}}, 0).
    \label{GSD-CM-S}
\end{equation}
Then, it would be evolved by
\begin{equation}
    i \hbar \partial_t \Psi(\mathbf{r}, \bar{\mathbf{r}}, t)
    =
    [W(\mathbf{r}, \bar{\mathbf{r}}, t) + U(|\mathbf{r} - \bar{\mathbf{r}}|)] \Psi(\mathbf{r}, \bar{\mathbf{r}}, t),
    \label{GSD-CM-E}
\end{equation}
where
\begin{equation}
    W(\mathbf{r}, \bar{\mathbf{r}}, t)
    \equiv
    H(\mathbf{r}, t) + H(\bar{\mathbf{r}}, t)
\end{equation}
and~$U(|\mathbf{r} - \bar{\mathbf{r}}|)$ is the gravitational interaction potential between the c.m. and its clone (regularized by a critical length~$L_\text{c} \sim L_\text{P}$):
\begin{equation}
    U(|\mathbf{r} - \bar{\mathbf{r}}|)
    \equiv
    - m^2 / |\mathbf{r} - \bar{\mathbf{r}}|_{L_\text{c}}
    \label{GSD-CM-IP}
\end{equation}
with
\begin{equation}
    |\mathbf{r} - \bar{\mathbf{r}}|_{L_\text{c}}
    \equiv
    \sqrt{|\mathbf{r} - \bar{\mathbf{r}}|^2 + L_\text{c}^2}.
\end{equation}
Note that the introduction of~$L_\text{c} \sim L_\text{P}$ in~$|\mathbf{r} - \bar{\mathbf{r}}|_{L_\text{c}}$ secures that it will never assume values smaller than the Planck length, where the very spacetime description (as a smooth manifold endowed with a Lorentzian metric) is expected to fail. Let us also note that Eqs.~\eqref{GSD-CM-S} and~\eqref{GSD-CM-E} ensure that the clone accurately imitates the behavior of the system's c.m. Ruled by Eq.~\eqref{GSD-CM-E}, the c.m. will, in general, lose information due to the interaction with the clone. This is to be interpreted as an effective loss of information to the (inaccessible) spacetime quantum degrees of freedom. We note that, in contrast to Caldeira-Leggett-like decoherence models, where the background environment privileges some reference frame, our mechanism does not, as all action takes place in the proper frame of the c.m. In Ref.~\cite{2025-aguiar}, we have analyzed the consequences of the model for systems free of external potentials. Eventually, it was obtained that the c.m. of systems with~$m \ll M_\text{P}$ effectively behaves in accordance with QM, while the c.m. of systems with~$m \sim M_\text{P}$ experiences classicalization.

\section{Gravitational self-decoherence: internal-vibration modes}
\label{GSD-IVM}

Now, let us consider a composed system with arbitrary mass~$m$ and internal degrees of freedom decomposed in terms of vibrational modes (decoupled from the c.m.). Let us focus on a single normal mode, with frequency~$\omega \lesssim \Omega_\text{P}$, (mass-weighted) normal coordinate~$Q$, and Hamiltonian
\begin{equation*}
    \mathcal{H}
    =
    - (\hbar^2/ 2) \partial_Q^2 + \omega^2 Q^2 / 2.
\end{equation*}
Since the effectiveness of our gravitational self-decoherence model scales with mass, let us take the energy scale~$\mu \equiv \hbar \omega$, set by the mode, to rewrite the Hamiltonian as
\begin{equation}
    \mathcal{H}(q)
    =
    - [\hbar^2 / (2 \mu)] \partial_q^2 + \mu \omega^2 q^2 / 2,
\end{equation}
where~$q \equiv Q / \mu^{1 / 2}$. According to QM, the preparation of the mode in a state~$\phi(q, 0)$ leads it to be described by a wavefunction~$\phi(q, t)$ satisfying
\begin{equation}
    i \hbar \partial_t \phi(q, t)
    =
    \mathcal{H}(q) \phi(q, t).
    \label{QM-IVM-E}
\end{equation}

In order to adapt the gravitational self-decoherence model~\cite{2025-aguiar} to handle the internal-vibration modes, we similarly assume that the normal mode with frequency~$\omega$ has a virtual clone with which it will be gravitationally coupled. As a result, by preparing the mode in a state~$\phi(q, 0)$, the clone would be driven to~$\phi(\bar{q}, 0)$, and the combined wavefunction
\begin{equation}
    \Phi(q, \bar{q}, 0)
    =
    \phi(q, 0) \phi(\bar{q}, 0)
    \label{GSD-IVM-S}
\end{equation}
would be evolved by
\begin{equation}
    i \hbar \partial_t \Phi(q, \bar{q}, t)
    =
    [\mathcal{W}(q, \bar{q}) + \mathcal{U}(|q - \bar{q}|)] \Phi(q, \bar{q}, t).
    \label{GSD-IVM-E}
\end{equation}
Here,
\begin{equation}
    \mathcal{W}(q, \bar{q})
    \equiv
    \mathcal{H}(q) + \mathcal{H}(\bar{q})
\end{equation}
and~$\mathcal{U}(|q - \bar{q}|)$ is the gravitational interaction potential between the mode and its clone:
\begin{equation}
    \mathcal{U}(|q - \bar{q}|)
    \equiv
    - \mu^2 / |q - \bar{q}|_{L_\text{c}}
    \label{GSD-IVM-IP}
\end{equation}
with
\begin{equation}
    |q - \bar{q}|_{L_\text{c}}
    \equiv
    \sqrt{|q - \bar{q}|^2 + L_\text{c}^2}.
\end{equation}
Similarly to the c.m. case, Eq.~\eqref{GSD-IVM-E} allows the clone to steal quantum coherence from the normal mode. Next, we show that such a mechanism is highly inefficient for normal modes with~$\omega \ll \Omega_\text{P}$, as those prepared in the laboratory.

To probe the effect of the interaction potential~\eqref{GSD-IVM-IP} in Eq.~\eqref{GSD-IVM-E}, we will use perturbation theory. Let us first look at the nonperturbed version of Eq.~\eqref{GSD-IVM-E}:
\begin{equation}
    i \hbar \partial_t \Theta(q, \bar{q}, t)
    =
    \mathcal{W}(q, \bar{q}) \Theta(q, \bar{q}, t).
    \label{GSD-IVM-E0}
\end{equation}
Let us write the corresponding energy eigenstates as
\begin{equation}
    \Theta_{n \bar{n}}(q, \bar{q})
    =
    \phi_n(q) \phi_{\bar{n}}(\bar{q})
    \quad
    (n, \bar{n} = 0, 1, 2, \ldots)
\end{equation}
and eigenvalues as
\begin{equation}
    \mathcal{W}_{n \bar{n}}
    =
    \mathcal{H}_n + \mathcal{H}_{\bar{n}}.
\end{equation}
Here,
\begin{equation}
    \phi_n(q)
    =
    \frac{1}{(2^n \, n!)^{1 / 2}} \frac{e^{- q^2 / (4 \beta^2)}}{(2 \pi \beta^2)^{1 / 4}} H_n\left(\frac{q}{\beta \sqrt{2}}\right)
\end{equation}
are the energy eigenstates of Eq.~\eqref{QM-IVM-E}, with~$H_n(w)$ being the Hermite polynomials and~$\beta \equiv \sqrt{\hbar / (2 \mu \omega)}$, and
\begin{equation}
    \mathcal{H}_n
    =
    \hbar \omega (n + 1 / 2)
\end{equation}
are the corresponding eigenenergies. In particular, the ground state of Eq.~\eqref{GSD-IVM-E0} is
\begin{equation}
    \Theta_{0 0}(q, \bar{q})
    =
    \phi_0(q) \phi_0(\bar{q})
\end{equation}
with energy
\begin{equation}
    \mathcal{W}_{0 0}
    =
    \hbar \omega,
\end{equation}
where
\begin{equation}
    \phi_0(q)
    =
    \frac{e^{- q^2 / (4 \beta^2)}}{(2 \pi \beta^2)^{1 / 4}}.
\end{equation}

To obtain the regime in which we are permitted to apply perturbation theory to evaluate the ground state of Eq.~\eqref{GSD-IVM-E}, let us calculate the mean value of the potential~$\mathcal{U}(|q - \bar{q}|)$ in the state~$\Theta_{0 0}(q , \bar{q})$:
\begin{eqnarray}
    \langle \mathcal{U} \rangle_{\Theta_{0 0}}
    &\equiv&
    \int\limits_{- \infty}^{+ \infty} \! d q \int\limits_{- \infty}^{+ \infty} \! d \bar{q} \; \Theta_{0 0}^*(q, \bar{q}) \mathcal{U}(|q - \bar{q}|) \Theta_{0 0}(q, \bar{q})
    \nonumber \\
    &=&
    - \frac{\mu^2 e^{L_\text{c}^2 / (8 \beta^2)} K_0(L_\text{c}^2 / (8 \beta^2))}{2 \beta \sqrt{\pi}},
\end{eqnarray}
where~$K_\alpha(w)$ are the modified Bessel functions of the second kind. By noticing that
\begin{equation}
    \left|\frac{\langle \mathcal{U} \rangle_{\Theta_{0 0}}}{\hbar \omega}\right|
    \ll
    1
\end{equation}
for~$\omega \ll \Omega_\text{P}$, we will use non-degenerate (first-order) perturbation theory to resolve Eq.~\eqref{GSD-IVM-E}, assuming normal modes with frequencies far from the Planck scale. As a result, we write the ground state as
\begin{equation}
    \Phi_{0 0}(q, \bar{q})
    =
    \sum\limits_{n, \bar{n} = 0}^\infty \Gamma_{n \bar{n}} \Theta_{n \bar{n}}(q, \bar{q}),
    \label{GSD-IVM-GS}
\end{equation}
with~$\Gamma_{n \bar{n}} \equiv \gamma_{n \bar{n}} / Z$,
\begin{equation}
    Z
    \equiv
    \left(\sum\limits_{n, \bar{n} = 0}^\infty |\gamma_{n \bar{n}}|^2\right)^{1 / 2},
\end{equation}
and~$\gamma_{n \bar{n}} = 1$ for~$n = \bar{n} = 0$, and
\begin{equation}
    \gamma_{n \bar{n}}
    =
    \int\limits_{- \infty}^{+ \infty} \! d q' \int\limits_{- \infty}^{+ \infty} \! d \bar{q}' \; \frac{\Theta_{n \bar{n}}^*(q', \bar{q}') \mathcal{U}(|q' - \bar{q}'|) \Theta_{0 0}(q', \bar{q}')}{\mathcal{W}_{0 0} - \mathcal{W}_{n \bar{n}}}
\end{equation}
otherwise. Correspondingly, the ground-state energy is 
\begin{equation}
    \mathcal{E}_{0 0}
    =
    \mathcal{W}_{0 0} + \langle \mathcal{U} \rangle_{\Theta_{0 0}}.
\end{equation}

\begin{figure}[t]
    \includegraphics[width = 80mm]{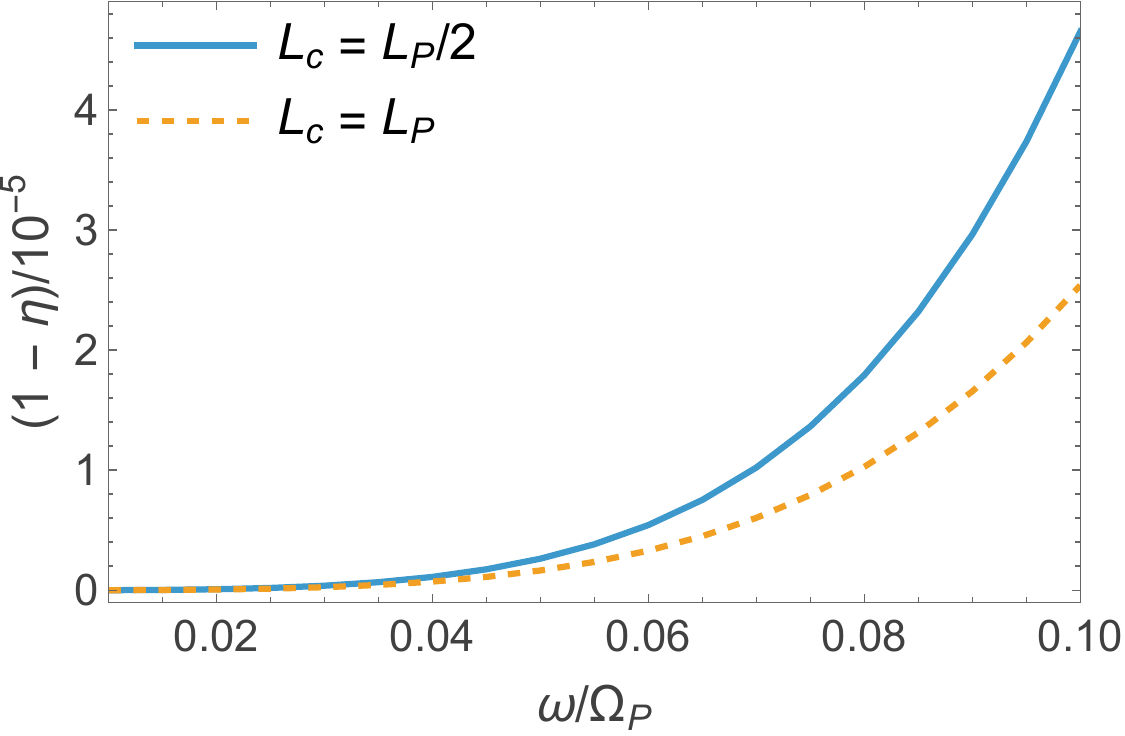}
    \caption{The plot exhibits how much a normal mode with frequency~$\omega$ fails to keep its purity~$\eta = 1$ in the ground state of the full system due to the corresponding interaction potential for two values of~$L_\text{c} \sim L_\text{P}$.}
    \label{fig_eta-critical}
\end{figure}

The physical picture that emerges from the discussion above is that, even in an ideal case where the full system is set in its ground state, the normal modes and their corresponding clones will ultimately become entangled due to the gravitational interaction. It rests with us to calculate how much coherence each mode would lose after tracing over the degrees of freedom of the corresponding clones, verifying that our model allows macroscopic systems to sustain the coherence of internal-vibration modes, as observed in recent experiments. For this purpose, let us trace over the clone degrees of freedom in the state~\eqref{GSD-IVM-GS}, obtaining the density-matrix elements
\begin{equation}
    \rho_{n n'}
    =
    \sum_{\bar{n} = 0}^\infty \Gamma_{n \bar{n}} \Gamma_{n' \bar{n}}^*
    \label{GSD-IVM-DM}
\end{equation}
of the normal mode, expressed in the basis~$\{\phi_n(q)\}_{n = 0}^\infty$. Thus, the corresponding purity of the mode is
\begin{eqnarray}
    \eta
    &=&
    \sum\limits_{n, n' = 0}^\infty |\rho_{n n'}|^2
    \nonumber \\
    &=&
    \sum\limits_{n, n', \bar{n}, \bar{n}' = 0}^\infty \Gamma_{n \bar{n}} \Gamma_{n' \bar{n}}^* \Gamma_{n' \bar{n}'} \Gamma_{n \bar{n}'}^*.
\end{eqnarray}
The larger the~$|\mathcal{U}(|q - \bar{q}|)|$, the larger the entanglement in~$\Phi_{0 0}(q, \bar{q})$, and the smaller the~$\eta$. Therefore, the larger the~$\omega$, the smaller the~$\eta$. Figure~\ref{fig_eta-critical} plots~$1 - \eta$ to show how much a normal mode with frequency~$\omega$ fails to keep its purity~$\eta = 1$ in the ground state of the full system due to the corresponding interaction potential. Let us note that, for all practical purposes,~$\omega \ll \Omega_\text{P}$, the purity of the mode is extremely close to unity.

\begin{figure}[t]
    \includegraphics[width = 80mm]{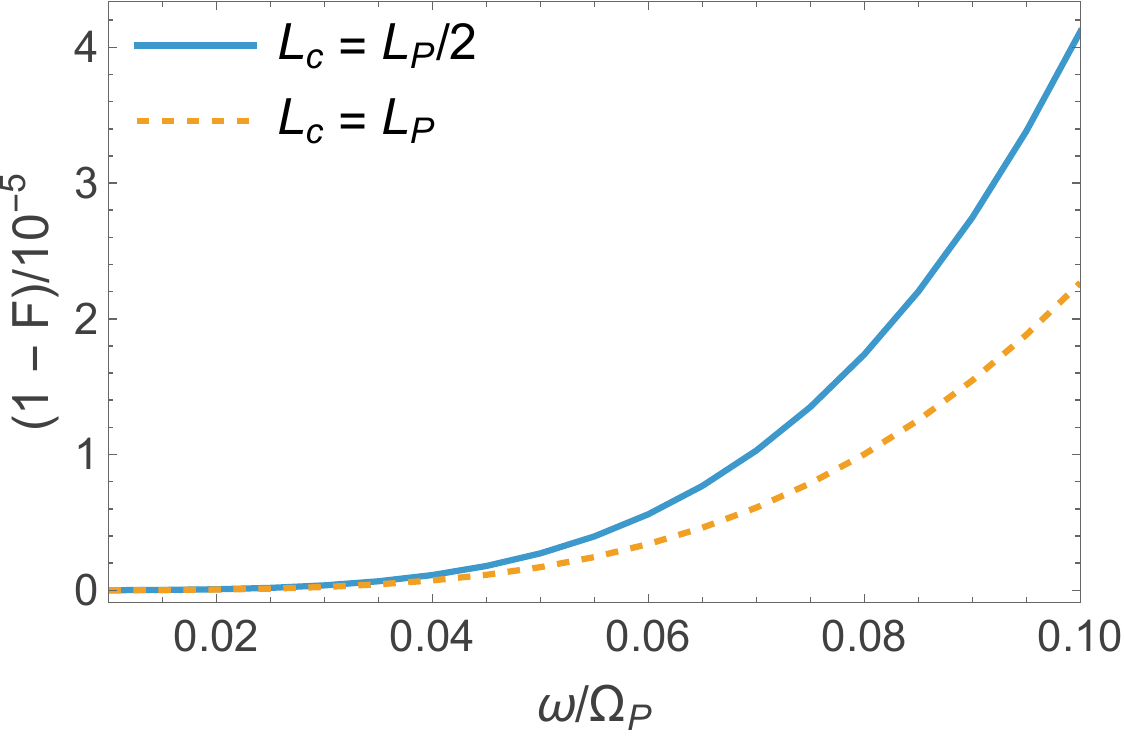}
    \caption{The graph plots how much the fidelity~$F$ between the state~\eqref{GSD-IVM-DM}, associated with a normal mode with frequency~$\omega$, and~$\phi_0(q)$ deviates from unity in the ground state of the full system due to the corresponding interaction potential for two values of~$L_\text{c} \sim L_\text{P}$.}
    \label{fig_F-critical}
\end{figure}

We can also calculate the fidelity
\begin{equation}
    F
    =
    \rho_{0 0}
    =
    \sum_{\bar{n} = 0}^\infty |\Gamma_{0 \bar{n}}|^2
\end{equation}
between the state~\eqref{GSD-IVM-DM}, associated with the normal mode, and~$\phi_0(q)$. The larger the entanglement in~$\Phi_{0 0}(q, \bar{q})$, the smaller the~$F$. Therefore, the larger the~$\omega$, the smaller the~$F$. Figure~\ref{fig_F-critical} plots~$1 - F$ to show how much a normal mode with frequency~$\omega$ deviates from~$\phi_0(q)$ in the ground state of the full system due to the corresponding interaction potential. Let us note that, for all practical purposes,~$\omega \ll \Omega_\text{P}$, the fidelity between the state associated with the mode and~$\phi_0(q)$ is extremely close to unity.

The results exhibited in Figs.~\ref{fig_eta-critical} and~\ref{fig_F-critical} strongly suggest that internal-vibration modes with~$\omega \ll \Omega_P$ should effectively behave in accordance with QM, no matter whether they are hosted by macroscopic systems, $m \gg M_\text{P}$, with c.m. obeying CM~\cite{2025-aguiar}. This is in line with all present experiments (to the best of our knowledge) as, {\em e.g.}, Ref.~\cite{2023-bild}, where a superposition of coherent states with~$\omega \sim 10^{- 33} \, \Omega_\text{P}$ is prepared in a substrate with~$m \sim M_\text{P}$.

\section{Conclusions}
\label{C}

It was argued in Ref.~\cite{2025-aguiar} that QM rules out the existence of bona-fide clocks capable of measuring space and time intervals of the order of (or smaller than) the Planck scale. Since bona-fide clocks are necessary to define and test relativistic spacetimes, this implies that our present description of spacetime (as smooth manifolds endowed with Lorentzian metrics) will require some (possibly radical) amendment at the Planck scale. It was claimed, thus, that wave equations would not be trustworthy to describe the c.m. of macroscopic systems with masses~$m \gg M_\text{P}$ (Compton wavelengths~$\lambdabar \ll L_\text{P}$), since at such scales the very spacetime on which these equations rest would not be reliable. The reasoning above led to the introduction of a gravitational self-decoherence model, which describes how the c.m. of systems with~$m \sim M_\text{P}$ would classicalize~\cite{2025-aguiar}.

In this paper, we have verified that our previous conclusions are in agreement with recent experiments~\cite{2023-bild}, according to which macroscopic systems are shown to effectively hold the quantum coherence of internal-vibration modes with frequencies~$\omega \ll \Omega_\text{P}$. We note that a necessary test for the proposal put forward in Ref.~\cite{2025-aguiar} requires demonstrating that the c.m. of isolated systems with~$m \gtrsim M_P$ cannot be put in spatial superposition. Although this is a paramount experimental challenge~\cite{2015-kovachy, 2019-fein, 2021-margalit}, the fast development of quantum technologies is a reason for optimism~\cite{2014-arndt, 2021-gasbarri, 2022-marshman, 2024-kremer, 2025-muffato, 2025-skakunenko, 2025-muretova, 2025-givon, 2025-feldman, 2025-benjaminov, 2025-liran}.

\acknowledgments

The authors thank Juan P{\^ e}gas and Felipe Portales for their feedback. G.H.S.A. was fully supported by the S{\~ a}o Paulo Research Foundation (FAPESP) under grant~2022/08424-3. G.E.A.M. was partially supported by the National Council for Scientific and Technological Development and FAPESP under grants~301508/2022-4 and~2022/10561-9, respectively.

\end{document}